\documentclass[journal,twoside]{IEEEtran}
\usepackage{amsmath,amssymb,amsfonts}
\usepackage{graphicx}
\usepackage{algorithm}
\usepackage{algorithmic}
\usepackage{textcomp}
\usepackage[hidelinks]{hyperref}
\usepackage{booktabs}
\usepackage{threeparttable}
\usepackage[figuresright]{rotating}
\usepackage{cite}
\usepackage{amsmath, amssymb, amsfonts, stmaryrd}

\def\BibTeX{{\rm B\kern-.05em{\sc i\kern-.025em b}\kern-.08em
    T\kern-.1667em\lower.7ex\hbox{E}\kern-.125emX}}
\begin{document}
\title{UT-OSANet: A Multimodal Deep Learning model for
Evaluating and Classifying Obstructive Sleep Apnea}
\author{Zijian Wang, Xiaoyu Bao, Chenhao Zhao, Jihui Zhang, Sizhi Ai, and Yuanqing Li \IEEEmembership{Fellow, IEEE}.
\thanks{This work was supported in part by the STI 2030--Major Projects, China under Grant 2022ZD0208900; in part by the Key Research and Development Program of Guangdong Province, China under Grant 2018B030339001.
(Corresponding author: Yuanqing Li and SiZhi Ai)}
\thanks{Yuanqing Li is with the School of Automation Science and Engineering, South China University of Technology, Guangzhou 510641, China, and also with the Brain Computer Intelligence Center, Artificial Intelligence and Digital Economy Laboratory, Guangzhou 510220, China (e-mail: auyqli@scut.edu.cn).}
\thanks{Sizhi Ai is with the Brain Computer Intelligence Center, Artificial Intelligence and Digital Economy Laboratory, Guangzhou 510220, China, and also with the Center for Sleep and Circadian Medicine, The Affiliated Brain Hospital, Guangzhou Medical University, Guangzhou 511436, China.}
\thanks{Zijian Wang is with the School of Automation Science and Engineering, South China University of Technology, Guangzhou 510641, China, and also with the Brain Computer Intelligence Center, Artificial Intelligence and Digital Economy Laboratory, Guangzhou 510220, China.}
\thanks{Xiaoyu Bao is with the School of Automation Science and Engineering, South China University of Technology, Guangzhou 510641, China, and also with the Brain Computer Intelligence Center, Artificial Intelligence and Digital Economy Laboratory, Guangzhou 510220, China.}
\thanks{Chenhao Zhao is with the School of Automation Science and Engineering, South China University of Technology, Guangzhou 510641, China, and also with the Brain Computer Intelligence Center, Artificial Intelligence and Digital Economy Laboratory, Guangzhou 510220, China.}
\thanks{Jihui Zhang is with the Center for Sleep and Circadian Medicine, The Affiliated Brain Hospital, Guangzhou Medical University, Guangzhou 511436, China.}
}
\maketitle

\begin{abstract}
Obstructive sleep apnea (OSA) is a highly prevalent sleep disorder that is associated with increased risks of cardiovascular morbidity and all-cause mortality. While existing diagnostic approaches can roughly classify OSA severity or detect isolated respiratory events, they lack the precision and comprehensiveness required for high-resolution, event-level diagnosis. Here, we present UT-OSANet, a deep learning–based model designed as a event-level, m, narihomeo diagnostic tool for OSA. 
This model facilitates detailed identification of events associated with OSA, including apnea, hypopnea, oxygen desaturation, and arousal. Moreover, the model employs flexibly adjustable input modalities such as electroencephalography (EEG), airflow, and $SpO_2$. It utilizes a random masked modality combination training strategy, allowing it to comprehend cross-modal relationships while sustaining consistent performance across varying modality conditions.
This model was trained and evaluated utilizing 9,021 polysomnography (PSG) recordings from five independent datasets. achieving sensitivities up to 0.93 and macro-F1 scores of 0.84–0.85 across research, clinical, and home scenarios. 
This model serves as an event-level, multi-scenario diagnostic instrument for real-world applications of OSA, while also establishing itself as a means to deepen the mechanistic comprehension of respiratory processes in sleep disorders and their extensive health implications.
\end{abstract}


\section{Introduction}
Obstructive sleep apnea (OSA) is a highly prevalent sleep disorder characterized by repeated upper airway collapse during sleep, leading to intermittent hypoxia and sleep fragmentation \cite{qureshi2003obstructive}. OSA is associated with various adverse clinical outcomes, including cardiovascular diseases such as hypertension, coronary artery disease, arrhythmia, and stroke, as the repeated incidents of nocturnal hypoxia result in the activation of the sympathetic nervous system and increased blood pressure fluctuations \cite{randerath2025central, bouloukaki2024preserved}. Additionally, OSA contributes to metabolic disorders, cognitive impairment, and an increased risk of neurodegenerative diseases such as Alzheimer's disease \cite{vanek2020obstructive, kerner2016obstructive}. OSA is also linked to various psychological conditions, including anxiety, depression, and emotional instability, with OSA patients exhibiting a higher prevalence of mental health disorders than observed among the general population \cite{kendzerska2024obstructive, bardwell2001response}. Moreover, excessive daytime sleepiness caused by OSA significantly reduces concentration, increases accident risk, and impairs overall quality of life\cite{javaheri2020update, panossian2012daytime, lal2021excessive}.
The prevalence of OSA varies across populations and affects both adults and children, with studies estimating that 1.2\% to 25\% of children aged 5 to 12 years are affected by OSA \cite{senaratna2017prevalence, ai2022blood}.
However, despite the high prevalence of OSA, accurate and accessible diagnostic methods are lacking.

Recently, promising machine learning and deep learning approaches have emerged as efficient and cost-effective alternatives to traditional polysomnography (PSG) for assessing OSA \cite{ferreira2022enabling, duarte2023role}. These methods often rely on a reduced set of physiological signals or focus on estimating coarse indicators such as the apnea–hypopnea index (AHI) \cite{thornton2012aasm}.
However, most existing algorithms face three major limitations. First, they are typically validated on relatively small and homogeneous cohorts, raising concerns about their generalizability to diverse populations. Second, current models primarily provide binary or severity-level classification and rarely perform fine-grained, event-level detection of OSA-related disturbances such as apnea, hypopnea, desaturation, and arousal. Third, they are usually designed for a single application scenario—either home-based screening or clinical evaluation—without a unified framework that can flexibly adapt across home, clinical, and research contexts \cite{hwang2017real, garde2015pulse, levy2023deep, berry2012respiratory, koley2013real, van2020portable, ben2012obstructive, huttunen2022comparison, yook2024deep, rossi2023sleep, zahid2023msed}.

Given the heterogeneous nature of OSA—with variations across age groups, comorbidities, and lifestyle factors \cite{senaratna2017prevalence, ai2022blood}—and the distinct demands of different usage scenarios \cite{park2024fda, thornton2012aasm, oksenberg2023duration}, achieving consistent and accurate diagnosis across settings remains a major challenge. In clinical environments, the availability and quality of physiological modalities differ markedly among institutions: while some employ full polysomnography (PSG) with comprehensive cardiorespiratory and neurophysiological monitoring, others rely on simplified channel configurations constrained by equipment, staffing, or cost. In contrast, home-based screening systems often capture only limited signals such as EEG, leading to incomplete physiological characterization. This disparity in signal availability across contexts underscores the need for diagnostic models that can flexibly adapt to varying input modalities while maintaining event-level accuracy. Such adaptable models would not only enhance diagnostic accessibility across diverse healthcare and home settings but also support scalable, personalized management of OSA in real-world practice.

This study presents UT-OSANet, a deep learning–based model designed for large-scale, event-level OSA assessment. By leveraging multimodal physiological signals—including electroencephalography (EEG), nasal airflow, and $SpO_2$—the model achieves precise detection of key OSA-related events such as apnea, hypopnea, arousal, and oxygen desaturation. With a flexible input configuration, UT-OSANet can be applied across diverse scenarios, ranging from home-based population screening to clinical severity evaluation and research-grade analyses. This adaptability, combined with validation on large-scale datasets, highlights its potential as a high-accuracy, scalable solution for comprehensive OSA monitoring in real-world settings.

\section{METHODS}
\subsection{Data}
\begin{figure*}[t]
\centering
\includegraphics[width=0.75\textwidth]{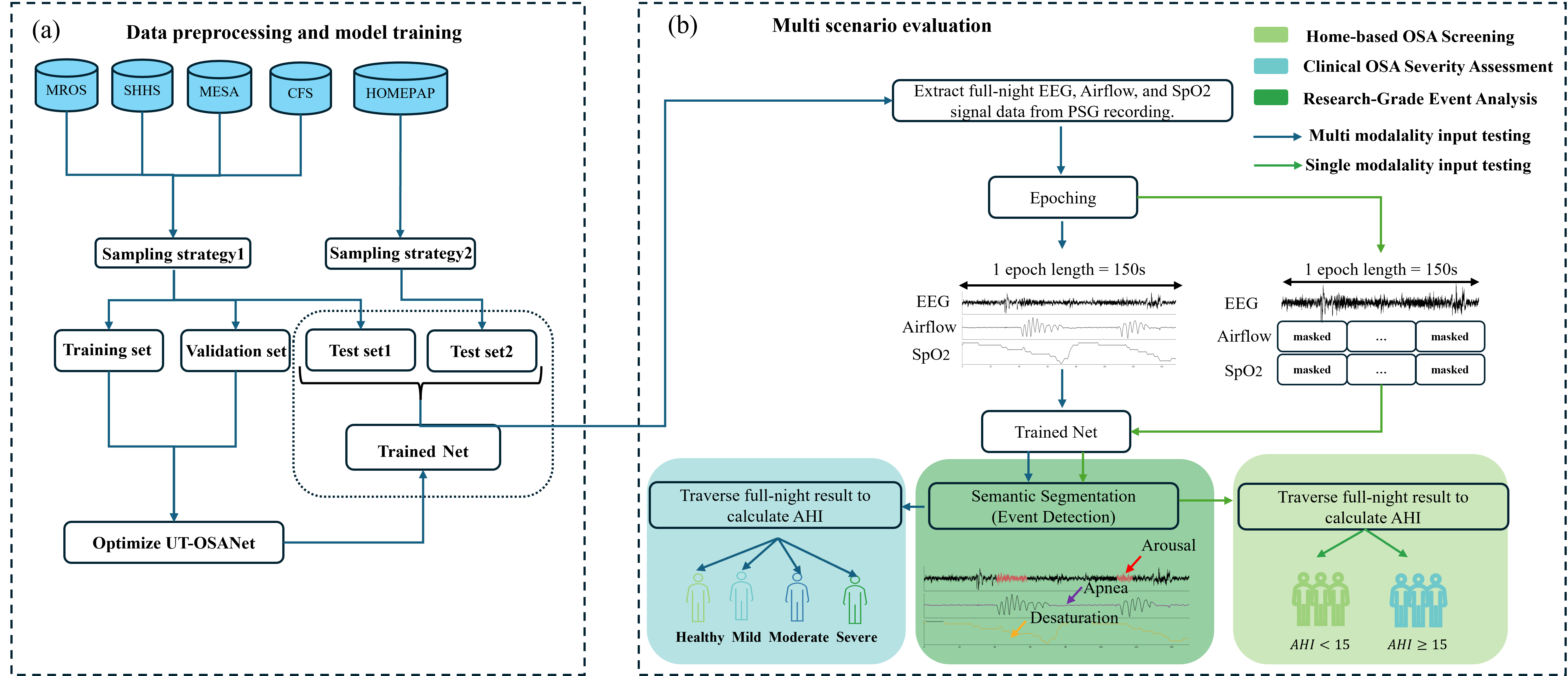}
\caption{Illustration of the study framework. (A) Data sources, sampling strategy, and model training process using polysomnography (PSG) recordings from multiple datasets (MROS, SHHS, MESA, CFS, and HOME-PAP). The dataset underwent preprocessing, excluding individuals with missing EDF files or severely distorted EEG, airflow, or $SpO_2$ signals. Two sampling strategies were employed: Sample Strategy 1 allocated 10\% of the remaining subjects to the test set, ensuring a balanced distribution across AHI categories ($AHI < 5$, 5$\leq AHI<15$, 15$\leq AHI<30$, and $AHI \geq 30$), while the remaining 90\% was split into a training set (75\%) and a validation set (15\%) through random sampling. Sample Strategy 2 excluded poor-quality recordings and used the entire dataset as the test set, allowing an evaluation of the model’s generalization performance. (B) Model application scenarios, including Home-based Moderate-to-Severe OSA Screening (classification of individuals with AHI < 15 or AHI $\geq$ 15), Clinical Severity Assessment (categorizing individuals into no OSA, mild, moderate, and severe OSA based on AHI thresholds), and Research-Grade Event Analysis (identification of apnea, hypopnea, desaturation, and arousal events).}\label{fig_frame}
\end{figure*}

A total of 9,021 PSG recordings from MROS, SHHS, MESA, and CFS were used for model development and validation. The HOMEPAP dataset was used solely as an independent test set to assess generalization. (Fig.~\ref{fig_frame}). As summarized in Table~\ref{tab1}, these datasets represent a diverse population with variations in age, sex, body mass index (BMI), and sleep characteristics. All recordings were acquired using laboratory-based PSG systems, with total sleep times (TSTs) ranging from 61.44 to 115.51 minutes across datasets. Sleep efficiency and wake after sleep onset (WASO) metrics further highlight inter-cohort variability in sleep quality. Apnea, hypopnea, $SpO_2$ desaturation, and arousal events were derived from publicly available expert annotations, and AHI was computed following the American Academy of Sleep Medicine (AASM) guidelines \cite{berry2017aasm}.

\textbf{Ethics statement.} All data used in this study were obtained from the National Sleep Research Resource (NSRR)\cite{zhang2018national}, which provides de-identified data collected under protocols approved by the respective institutional review boards of the original studies. Written informed consent was obtained from all participants by the original investigators. As this study involves only secondary analyses of existing, de-identified datasets, additional ethical approval was not required.

To analyze OSA-related events, we examined the distributions of apnea, hypopnea, arousal, and $SpO_2$ desaturation events across datasets, as detailed in Table~\ref{tab2}. The SHHS dataset included the most events (2,212,230), whereas the HOMEPAP dataset included the fewest events (74,882). The high prevalence of hypopnea events in the SHHS and MESA datasets suggests dataset-specific variations in composition and scoring, underscoring the need for cross-dataset validation to ensure the generalizability of the developed model.

Following preprocessing, exclusion criteria were applied to remove recordings with missing AHI values, severe signal noise in the EEG, $SpO_2$, or airflow channel data, and incomplete sleep studies. The final dataset was split into training (75\%), validation (15\%), and test (10\%) sets, and the HOMEPAP dataset was retained as an independent test set for external evaluation.

\begin{table*}[t]
\centering
\caption{Summary metrics of polysomnography data used for training and testing the models}
\label{tab1}
\begin{threeparttable}
\begin{tabular}{lccccc}
\toprule
 & \textbf{MROS} & \textbf{SHHS} & \textbf{MESA} & \textbf{CFS} & \textbf{HOMEPAP} \\
\midrule
N             & 2900           & 4940           & 940            & 660            & 131            \\
N (train)     & 2179           & 3705           & 705            & 495            & --             \\
N (val)       & 435            & 741            & 141            & 99             & --             \\
N (test)      & 290            & 494            & 94             & 66             & 131            \\
Age [years]   & 76.36 $\pm$ 5.47   & 63.13 $\pm$ 11.22  & 69.63 $\pm$ 9.21   & 41.40 $\pm$ 19.34  & 63.13 $\pm$ 11.22  \\
Sex (\% male) & 100\%    & 76\%    & 23\%    & 23\%    & 76\%    \\
BMI [kg/m$^2$] & 27.18 $\pm$ 3.81   & 28.16 $\pm$ 5.09   & 28.72 $\pm$ 5.60   & 32.36 $\pm$ 9.54   & 28.16 $\pm$ 5.09   \\
TST [min]     & 115.51 $\pm$ 66.73 & 61.44 $\pm$ 44.03  & 94.51 $\pm$ 65.35  & 82.19 $\pm$ 65.75  & 61.44 $\pm$ 44.03  \\
WASO [min]    & 355.51 $\pm$ 69.41 & 359.83 $\pm$ 64.56 & 359.80 $\pm$ 82.60 & 372.82 $\pm$ 75.28 & 359.83 $\pm$ 64.56 \\
SE            & 76.04 $\pm$ 12.03  & 82.77 $\pm$ 10.55  & 75.74 $\pm$ 13.46  & 78.67 $\pm$ 12.96  & 82.77 $\pm$ 10.55  \\
NREM 1 [\%]   & 6.86 $\pm$ 4.35    & 5.44 $\pm$ 3.95    & 14.39 $\pm$ 9.21   & 5.04 $\pm$ 3.87    & 5.44 $\pm$ 3.95    \\
NREM 2 [\%]   & 62.77 $\pm$ 9.68   & 56.54 $\pm$ 11.72  & 57.56 $\pm$ 10.26  & 56.08 $\pm$ 12.88  & 56.54 $\pm$ 11.72  \\
NREM 3 [\%]   & 11.25 $\pm$ 9.03   & 18.21 $\pm$ 11.87  & 10.04 $\pm$ 9.01   & 20.59 $\pm$ 13.43  & 18.21 $\pm$ 11.87  \\
REM [\%]      & 19.25 $\pm$ 6.66   & 19.80 $\pm$ 6.27   & 18.02 $\pm$ 6.69   & 18.16 $\pm$ 7.27   & 19.80 $\pm$ 6.27   \\
ArI [h$^{-1}$] & 23.64 $\pm$ 11.73  & 19.16 $\pm$ 10.66  & 22.32 $\pm$ 12.06  & 15.54 $\pm$ 10.00  & 19.16 $\pm$ 10.66  \\
AHI [h$^{-1}$] & 21.35 $\pm$ 16.30  & 17.94 $\pm$ 16.11  & 24.15 $\pm$ 19.55  & 12.53 $\pm$ 17.01  & 17.94 $\pm$ 16.11  \\
PLMI [h$^{-1}$] & 35.72 $\pm$ 37.53  & --                 & 14.21 $\pm$ 24.98  & 8.05 $\pm$ 17.94   & --                 \\
$SpO_2<80$ [min] & 31.83 $\pm$ 190.03 & 67.58 $\pm$ 767.16 & 96.81 $\pm$ 658.13 & 22.53 $\pm$ 169.02 & 67.58 $\pm$ 767.16 \\
\bottomrule
\end{tabular}
\begin{tablenotes}
\footnotesize
\item Data are presented as mean $\pm$ standard deviation or count (n). BMI, body mass index; WASO, wake after sleep onset; TST, total sleep time; SE, sleep efficiency; AHI, apnea-hypopnea index; ArI, arousal index; REM, percentage of total sleep time in REM sleep; NREM 1/2/3, stage 1/2/3 percentages; PLMI, periodic limb movement index; $SpO_2<80$, minutes with oxygen saturation $<80\%$.
\end{tablenotes}
\end{threeparttable}
\end{table*}

\begin{table*}[t]
\centering
\caption{Number of detected OSA-related events across datasets}
\label{tab2}
\begin{threeparttable}
\begin{tabular}{lccccc}
\toprule
\textbf{Dataset} & \textbf{Apnea} & \textbf{Hypopnea} & \textbf{Arousal} & \textbf{$SpO_2$ Desaturation} & \textbf{Total} \\
\midrule
MROS    & 68{,}485   & 118{,}447  & 254{,}040  & 574{,}773   & 1{,}015{,}745 \\
SHHS    & 140{,}463  & 891{,}332  & 556{,}762  & 623{,}673   & 2{,}212{,}230 \\
MESA    & 43{,}062   & 187{,}612  & 263{,}881  & 554{,}290   & 1{,}048{,}845 \\
CFS     & 14{,}427   & 61{,}878   & 67{,}781   & 56{,}062    & 200{,}148 \\
HOMEPAP & 3{,}274    & 14{,}083   & 20{,}559   & 36{,}966    & 74{,}882 \\
\bottomrule
\end{tabular}
\begin{tablenotes}
\item Apnea: apnea event count; Hypopnea: hypopnea event count; Arousal: arousal event count; $SpO_2$ desaturation: oxygen desaturation event count; Total: sum of all detected events per dataset.  
\end{tablenotes}
\end{threeparttable}
\end{table*}

\subsection{Scoring Rules}
Following the American Academy of Sleep Medicine (AASM) scoring manual \cite{thornton2012aasm} and the International Classification of Sleep Disorders, Third Edition (ICSD-3) guidelines \cite{adams2016icsd}, all respiratory and arousal events were defined as follows:
\begin{itemize}
    \item \textbf{Apnea:}
    A drop in peak airflow signal amplitude by $\geq 90\%$ of the pre-event baseline, measured via an oronasal thermal sensor (for diagnostic studies), PAP device flow (for titration studies), or an equivalent airflow sensor. 
    The event must last for at least 10~s.
    
    \item \textbf{Hypopnea:}
    A reduction in airflow signal amplitude by $\geq 30\%$ from baseline for $\geq 10$~s, accompanied by either a $\geq 3\%$ oxygen desaturation or a cortical arousal.

    \item \textbf{SpO$_2$ Desaturation:}
    A drop in oxygen saturation ($SpO_2$) of $\geq 3\%$ from the pre-event baseline, typically sustained for $\geq 3$~s and often following an apnea or hypopnea event.

    \item \textbf{Arousal:}
    An abrupt shift in EEG frequency (alpha, theta, or $>16$~Hz; spindles excluded) lasting $\geq 3$~s, following at least 10~s of stable sleep.
\end{itemize}

The \textit{apnea–hypopnea index (AHI)} was computed as:
\begin{equation}
    AHI = \frac{N_{\text{apnea}} + N_{\text{hypopnea}}}{T_{\text{sleep}}/60},
    \label{eq:ahi}
\end{equation}
where $N_{\text{apnea}}$ and $N_{\text{hypopnea}}$ denote the total counts of apnea and hypopnea events, and $T_{\text{sleep}}$ is the total sleep time in minutes.

\subsection{Event Morphology and Modalities}
\begin{figure*}[t]
    \centering
    \includegraphics[width=0.75\linewidth]{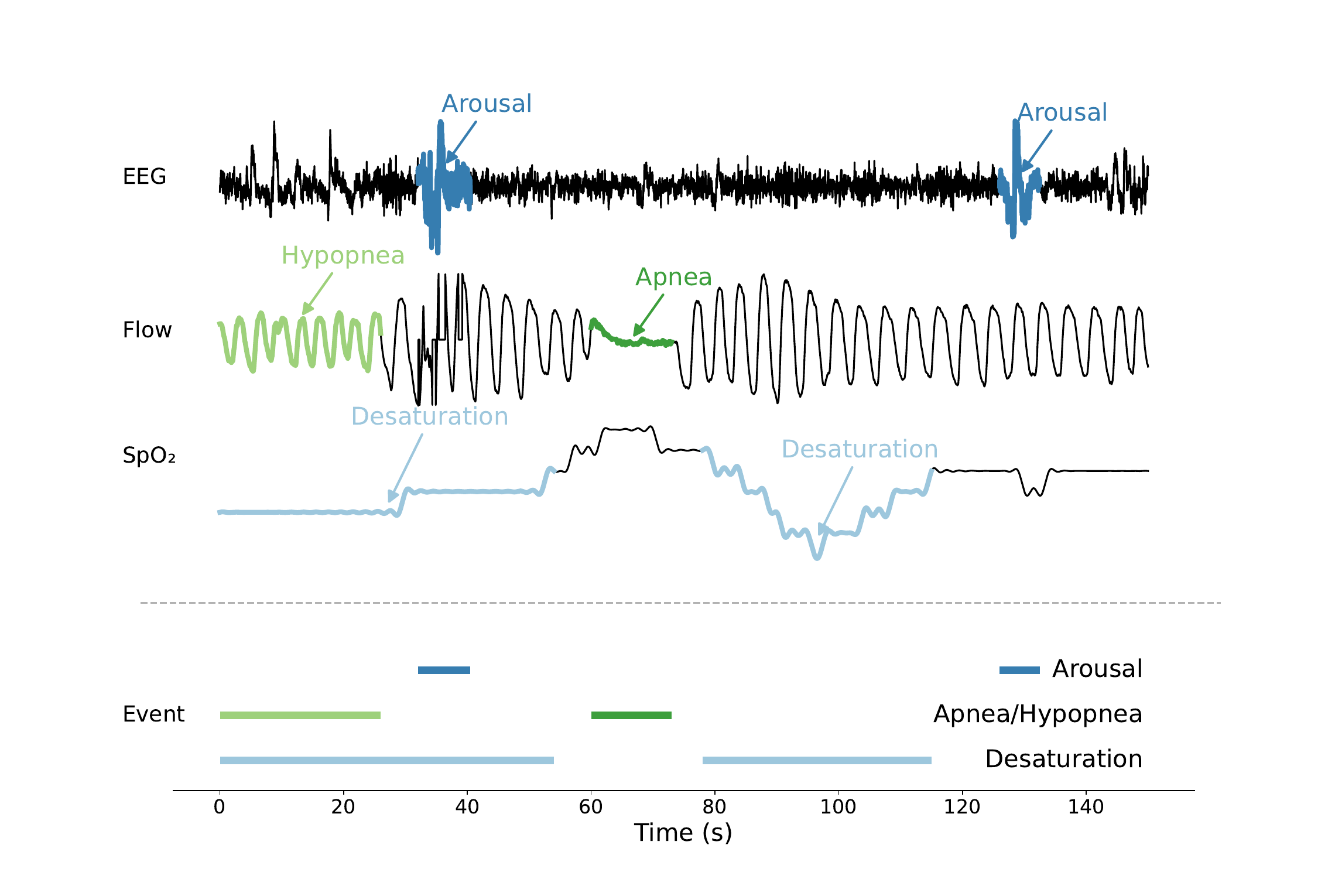}
    \caption{
The figure shows three synchronized signals: EEG (top), airflow (middle), and SpO$_2$ (bottom).
Apnea is marked by a complete loss of airflow, often followed by a delayed fall in SpO$_2$.
Hypopnea is shown as a partial reduction of airflow with a smaller change in SpO$_2$ or a brief EEG arousal.
EEG arousal appears as a short burst of faster activity after a respiratory disturbance.
An event bar is placed below the signals to mark the timing of arousal, apnea/hypopnea, and desaturation, allowing their order and overlap to be viewed in the same time window.
    }
    \label{fig_event}
\end{figure*}

To illustrate the temporal morphology and multimodal signatures of these events, Fig.~\ref{fig_event} shows representative PSG segments containing apnea, hypopnea, SpO$_2$ desaturation, and arousal events across EEG, airflow, and SpO$_2$ channels. Across these examples, the events follow a clear time pattern. A drop or loss of airflow comes first. The fall in SpO$_2$ appears after a short delay, as oxygen levels change more slowly than airflow. An EEG arousal often comes later in the same sequence, showing a brief increase in fast activity when the subject resumes breathing.
By placing EEG, airflow, and SpO$_2$ in the same window, the figure shows how these events line up in time and how one event is linked to the next.

\begin{itemize}
    \item \textbf{Apnea (Flow $\downarrow\!\downarrow$; EEG $\rightarrow$; SpO$_2$ delayed $\downarrow$):}
    Near-complete cessation of airflow for $\geq 10$~s, producing a flat or plateaued flow trace. EEG remains stable unless followed by an arousal, and SpO$_2$ exhibits a delayed decrement.

    \item \textbf{Hypopnea (Flow $\downarrow$; EEG $\rightarrow$/$\uparrow$; SpO$_2$ $\downarrow$ or arousal):}
    Partial reduction of airflow amplitude with preserved respiratory periodicity, typically accompanied by a mild desaturation ($\geq 3\%$) and/or EEG arousal.

    \item \textbf{SpO$_2$ Desaturation (SpO$_2$ $\downarrow$; Flow reduced earlier):}
    A gradual or step-like decline in $SpO_2$ following respiratory restriction, recovering after event termination.

    \item \textbf{Arousal (EEG burst; transient Flow/EMG increase):}
    A brief ($\geq 3$~s) EEG frequency shift showing fast activity (alpha, theta, or $>16$~Hz, excluding spindles), often accompanied by transient changes in airflow or muscle tone.
\end{itemize}

\subsection{Preprocessing and Exclusion Criteria}
To ensure data consistency and quality across all recordings, EEG, $SpO_2$, and airflow signals underwent standardized preprocessing steps (Fig.~\ref{fig_frame}). All signals were resampled to 100~Hz to achieve a uniform sampling rate across modalities, followed by Z-score normalization to standardize amplitude distributions\cite{zhang2018national}. EEG signals were further bandpass filtered between 0.5--45~Hz to remove noise and artifacts while preserving the frequency range relevant for sleep analysis\cite{vallat2021open}. Quality control was performed using Z-score outlier detection to identify and exclude extreme deviations. Participants were excluded if any of the following applied\cite{liu2021large}:  
1) missing EDF files,  
2) incomplete or missing AHI information,  
3) excessive noise or artifacts in EEG, $SpO_2$, or airflow channels that could not be adequately corrected.  
These preprocessing and exclusion procedures ensured that only high-quality, complete recordings were retained for subsequent model training, validation, and evaluation.

\subsection{UT-OSANet}
\subsubsection{Notation}
We denote by $[a,b]$ the set of integers $\{n \in \mathbb{N} \mid a \leq n \leq b\}$, 
with $\llbracket N \rrbracket$ being shorthand for $[1,N]$, 
and by $n \in \llbracket N \rrbracket$ the $n$-th sample in $\llbracket N \rrbracket$. 
A segment of multimodal physiological data is denoted by 
$\mathbf{x} \in \mathbb{R}^{C \times T}$, 
where $C$ is the number of available modalities (e.g., EEG, airflow, and $SpO_2$) 
and $T$ is the duration of the segment in samples. 

A binary modality mask is defined as 
$\mathbf{m} \in \{0,1\}^C$, 
where $m_c = 1$ indicates that the $c$-th modality is present and $m_c = 0$ otherwise. 

An event type is defined as 
$\varepsilon_i = (\varrho_i, \delta_i, l_i) \in \mathbb{R}_+^2 \times \mathcal{L}$, 
where $\varrho_i$, $\delta_i$, and $l_i$ denote the center time, duration, 
and label of the $i$-th event, respectively, and 
$\mathcal{L} = [L]$ represents the event label space. 
The set of $N_t$ true events within a given segment is denoted by 
$\varepsilon^t = \{\varepsilon_i^t \mid i \in \llbracket N_t \rrbracket\}$. 

The complete training sample is represented as 
$\chi = \{\mathbf{x}, \mathbf{m}, \varepsilon^t\}$, 
and we denote by $\chi \in \mathcal{D}_*$ a sample belonging to one of the training, 
validation, or testing subsets. 
For brevity, the batch dimension is omitted in the following model description.
In contrast to unimodal sleep analysis networks, UT-OSANet is a multimodal deep learning model designed to detect and classify OSA-related events (apnea, hypopnea, arousal, and desaturation) and to estimate the apnea–hypopnea index (AHI) (Fig.~\ref{fig_event}). The model takes as input a flexible combination of synchronized physiological modalities—including EEG, airflow, and $SpO_2$—which can be selectively included or omitted as available, and are preprocessed into overnight time-series segments. The input to the model was constructed as multimodal epochs of 250 s, each containing synchronized EEG, airflow, and $SpO_2$ signals. To enhance adaptability, any modality could be randomly masked or omitted during training, allowing the network to handle incomplete or variable input combinations while maintaining temporal consistency across datasets.

\subsubsection{Model overview}
Given an input set 
\[
\chi = \{\mathbf{x}, \mathbf{m}, \varepsilon^t\} 
\in \mathbb{R}^{C\times T} \times \{0,1\}^C \times \mathbb{R}_+^{N_t\times 2} \times \mathcal{L},
\]
containing multimodal physiological data $\mathbf{x}$ (e.g., EEG, airflow, and $SpO_2$)
with $C$ channels and $T$ time steps, a modality mask $\mathbf{m}$ indicating the 
availability of each channel, and the set of true events $\varepsilon^t$, 
the goal of the model is to detect all possible respiratory-related events within 
the segment. In this context, detection involves both the \textit{classification} 
and \textit{temporal localization} of each event in the signal space.
During training, any subset of modalities can be randomly masked according to 
$\mathbf{m}$\cite{leidal2017learning}, allowing the network to learn representations that remain robust 
to missing or incomplete modalities. The model generates a set of 

\textit{default event windows} 
\[
\varepsilon^d = \{\varepsilon^d_j \mid j \in \llbracket N_d \rrbracket\}
\]
for each input segment and matches each true event to its nearest default window 
if their intersection-over-union (IoU) is at least 0.2. At test time, the network outputs predicted event windows with class probabilities and temporal coordinates. A non-maximum suppression (NMS) procedure is then applied to remove redundant predictions that highly overlap (IoU $>$ 0.5)\cite{zheng2020distance} with higher-probability events. The final output of the model is the set of predicted events 
\[
\varepsilon^p = \{\mathbf{p}, \mathbf{y}\},
\]
where $\mathbf{p}$ denotes the predicted temporal parameters (onset and duration) 
and $\mathbf{y}$ the corresponding class probabilities.

\subsubsection{Network architecture}
The proposed network, termed \textbf{UT-OSA}, consists of two major components: 
a Cross-Modality U-Net and a Transformer Encoder, designed to capture both local 
hierarchical patterns and long-range temporal dependencies in multimodal data. 
The overall architecture jointly processes synchronized EEG, airflow, and $SpO_2$ signals to detect respiratory-related events (apnea, hypopnea, arousal, and desaturation).

\textbf{Cross-Modality U-Net}. This module serves as the backbone for local feature extraction and early multimodal fusion. Given an input segment $\mathbf{x} \in \mathbb{R}^{C\times T}$, where $C$ is the number of modalities and $T$ is the temporal length, the encoder extracts hierarchical features through a series of 
1D convolutional layers (kernel size = 3), each followed by batch normalization, ReLU activation, and max-pooling (stride = 2)\cite{ronneberger2015u}. Skip connections are maintained between encoder and decoder blocks to preserve fine-grained temporal information\cite{targ2016resnet}. Each encoder block captures modality-specific representations, while inter-stream interactions within the U-Net structure enable early cross-modality fusion\cite{gao2020survey}. The decoder progressively upsamples and integrates features through transposed convolutions and bottleneck blocks, producing a unified temporal representation $V_{\text{U-Net}}$ that encodes both spatial–temporal and cross-modal dependencies. Importantly, the U-Net maintains the original temporal alignment of multimodal inputs without disrupting sequence order, ensuring consistent temporal continuity for downstream modeling.

\textbf{Transformer encoder.} To capture long-range contextual relationships, the Transformer encoder processes $V_{\text{U-Net}}$ through multiple layers of multi-head self-attention (MHSA) and feed-forward sub-layers, each followed by residual ``Add \& Norm'' operations\cite{vaswani2017attention}. Positional encoding $\phi_{\text{pos}}$ is applied before attention computation to preserve temporal ordering:
\[
\tilde{V}_{\text{U-Net}} = \phi_{\text{pos}}(V_{\text{U-Net}}), \quad 
V_{\text{Trans}} = \phi_{\text{Trans}}(\tilde{V}_{\text{U-Net}}).
\]
Functionally, the Transformer block fulfills two complementary roles: 
(1) modeling long-range temporal dependencies within each modality, integrating information 
across distant time points to represent sleep–respiration dynamics; and 
(2) learning cross-modal event coupling, capturing how concurrent changes in EEG, airflow, 
and $SpO_2$ jointly shape apnea–hypopnea patterns. 
The resulting high-level feature vector $V_{\text{Trans}}$ encodes both intra- and inter-modality 
temporal dependencies for downstream fusion and classification.

\textbf{Fusion and output.} The final fusion feature is obtained by concatenating $V_{\text{U-Net}}$ and $V_{\text{Trans}}$, which is then passed through fully connected layers for event-level classification\cite{gao2021utnet}. The network outputs four probability maps corresponding to apnea, hypopnea, arousal, and desaturation 
classes, using a sigmoid activation:
\[
\mathbf{y} = \sigma(W[V_{\text{U-Net}}, V_{\text{Trans}}] + b), 
\quad \mathbf{y} \in \mathbb{R}^{T'\times 4}.
\]

\subsubsection{Loss Function}
To handle multilabel supervision and overlapping respiratory events, 
the network was trained using a weighted binary cross-entropy with logits loss:
\[
\mathcal{L}_{\text{BCE}} 
= -\frac{1}{N}\sum_{i=1}^{N}\sum_{k=1}^{K}
    w_k \big[
    y_{i,k} \log \sigma(\hat{y}_{i,k}) + 
    (1-y_{i,k}) \log (1-\sigma(\hat{y}_{i,k}))
    \big],
\]
where $\hat{y}_{i,k}$ and $y_{i,k}$ denote the predicted and true labels for the $k$-th class, 
$\sigma(\cdot)$ is the sigmoid activation, and $w_k$ is a class-specific weight 
used to mitigate class imbalance across apnea, hypopnea, arousal, and desaturation events. This corresponds to the PyTorch implementation \texttt{nn.BCEWithLogitsLoss},  which fuses the sigmoid and binary cross-entropy computations for numerical stability.  To ensure temporal smoothness and prevent over-fragmentation of detected events, 
an auxiliary continuity regularization term was introduced:
\[
\mathcal{L}_{\text{smooth}} 
= \frac{1}{N(T'-1)} \sum_{i=1}^{N}\sum_{t=1}^{T'-1}
    \|\hat{y}_{i,t+1} - \hat{y}_{i,t}\|_1,
\]
which penalizes abrupt changes between adjacent time steps in the predicted sequence.  
The total objective function is defined as
\[
\mathcal{L}_{\text{total}} 
= \mathcal{L}_{\text{BCE}} + \lambda \mathcal{L}_{\text{smooth}},
\]
where $\lambda = 0.1$ empirically balances classification accuracy and temporal stability.

\subsubsection{Training Strategy}
\textit{Optimization and regularization.}
All models were trained using the Adam optimizer with an initial learning rate of 
$1\times10^{-4}$, $\beta_1=0.9$, $\beta_2=0.999$, and a batch size of 8. 
A cosine annealing schedule gradually decayed the learning rate to $1\times10^{-6}$ over 50 epochs. 
Weight decay ($1\times10^{-5}$) was applied to prevent overfitting. 
Batch normalization, residual connections, and dropout layers (rate = 0.2–0.3) 
further improved training stability and generalization. 
Gradient clipping with a maximum norm of 5 was used to prevent gradient explosion.

\textit{Modality dropout.}
To enhance robustness against missing or noisy signals, 
a modality dropout strategy was applied during training. 
In each batch, one or more modalities (EEG, airflow, or $SpO_2$) were randomly masked 
with probability $p=0.3$. 
This strategy encouraged the model to learn cross-modal redundancy and maintain reliable predictions 
under incomplete multimodal configurations, mimicking real-world home-based conditions.


\begin{figure*}[t]
    \centering
    \includegraphics[width=0.75\linewidth]{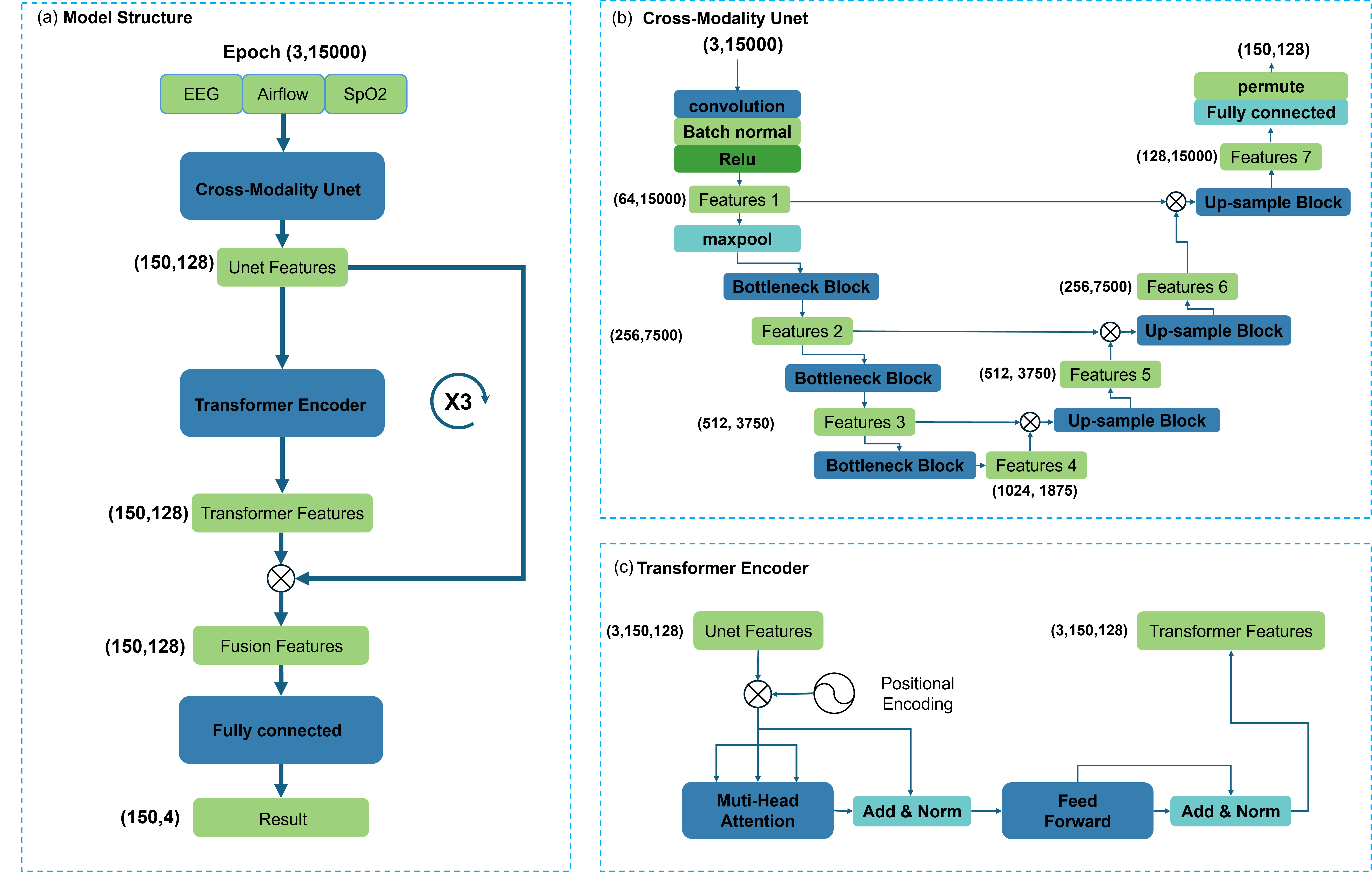}
    \caption{Model structure. The model integrates U-Net and transformer architectures. The cross-modality U-Net extracts features from EEG, airflow, and $SpO_2$ signals. Positional encoding enhances temporal information. The transformer learns long-range dependencies and fuses with U-Net features. The classifier predicts events such as apnea, hypopnea, arousal, and desaturation.}
    \label{fig_model_structure}
\end{figure*}

\subsection{Baseline Models}
To benchmark the performance of the proposed multimodal architecture, 
two baseline models were constructed. 
All models shared identical preprocessing, network structure, optimization settings, 
and evaluation metrics as the main model (Fig.~\ref{fig_frame}), 
but differed in the input modality configuration through controlled modality masking.

\textbf{Baseline Model 1: EEG-Only.}  
This model used only EEG signals as input, with $SpO_2$ and airflow channels masked out. 
It was designed to assess the discriminative capability of EEG alone for detecting and classifying 
obstructive sleep apnea (OSA)–related events, including apnea, hypopnea, and arousal.

\textbf{Baseline Model 2: $SpO_2$ + Airflow.}  
This model utilized only $SpO_2$ and airflow signals while masking EEG input. 
It enabled evaluation of the relative contribution of respiratory and oxygen saturation dynamics 
to the detection of respiratory events, particularly apnea, hypopnea, and desaturation.
Both baseline models followed the same data preprocessing pipeline 
(resampling, Z-score normalization, and artifact rejection), 
network configuration (layers, optimizers, and loss functions), 
and evaluation metrics (accuracy, precision, recall, and macro-F1). 
Performance comparison among the two baselines and the full multimodal model 
quantified the individual and joint contributions of EEG, $SpO_2$, and airflow modalities 
to overall detection accuracy.

\section{RESULTS}
\subsection{Research-Grade Event Analysis}
UT-OSANet was evaluated for event-level detection of apnea, hypopnea, arousal, and $SpO_2$ desaturation. Balanced F1-scores of 0.83–0.87 were obtained across datasets (Supplementary Table~S3). 

\begin{figure*}[t]
    \centering
    \includegraphics[width=0.75\textwidth]{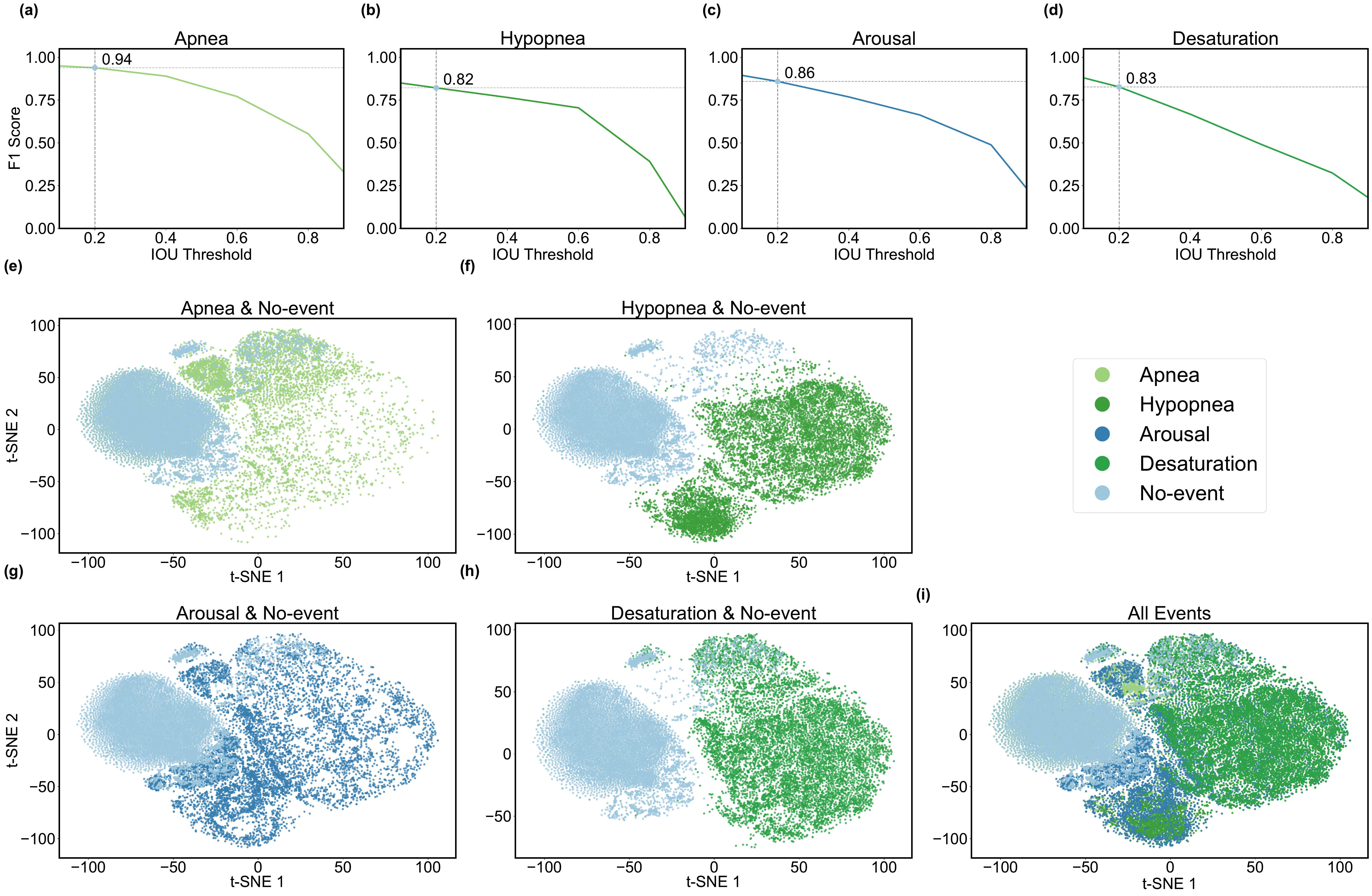}
    \caption{(a-e): t-SNE visualization of the model output, comparison of apnea, hypopnea, SpO2 desaturation, and arousal events and no OSA-related events; (f-i) F1 scores for event detection as a function of the IOU threshold, with the F1 score at IOU=0.2 serving as the final evaluation result. (a) to (d) present t-SNE projections of different OSA-related events compared with those of normal epochs. The F1 score variation curves (f--i) demonstrate how the detection performance for different OSA-related events changes as the IOU threshold increases. At IOU = 0.2, the model achieves its highest F1 score, with values of approximately 0.82 for apnea events, 0.86 for hypopnea events, 0.85 for arousal events, and 0.82 for SpO2 desaturation events.}
    \label{fig_tsne}
\end{figure*}

Figure \ref{fig_tsne} shows the event-wise predictions and feature space distributions. Specifically, Figure \ref {fig_tsne}(a--d)presents the variation in the F1 score for different intersection-over-union (IOU) thresholds. The model achieves the highest F1 score at an IOU of 0.2, with F1 scores of approximately 0.82 for apnea events, 0.86 for hypopnea events, 0.85 for arousal events, and 0.82 for $SpO_2$ desaturation events. As the IOU threshold increases, the F1 score gradually decreases, reflecting stricter event-matching criteria. For example, at an IOU of 0.5, the F1 score for apnea detection decreases to approximately 0.6, whereas the F1 scores for hypopnea and arousal detection remain higher, indicating more consistent prediction results for these types of events. Similar results are obtained for $SpO_2$ desaturation detection, reinforcing the model's robustness across different event types. The overall decrease in the F1 score at higher IOU thresholds highlights a trade-off between detection sensitivity and localization accuracy, suggesting that while the model can effectively identify events, precise temporal alignment remains challenging when stricter criteria are employed.

In addition to feature visualization, Figure~\ref{fig_tsne}(f--i) shows t-distributed stochastic neighbor embedding (t-SNE) projections, which display how apnea, hypopnea, arousal, and $SpO_2$ desaturation samples spread in the learned feature space relative to normal epochs. While each event forms its own cluster to some degree, Figure~\ref{fig_tsne}(i) also shows that samples from apnea, hypopnea, and $SpO_2$ desaturation partly overlap. This overlap is expected, as these events often occur in the same breathing cycle and share related signal changes. The mixed regions suggest that the model captures their close timing and the way these events influence each other.

\begin{figure*}[t]
\centering
\includegraphics[width=0.75\textwidth]{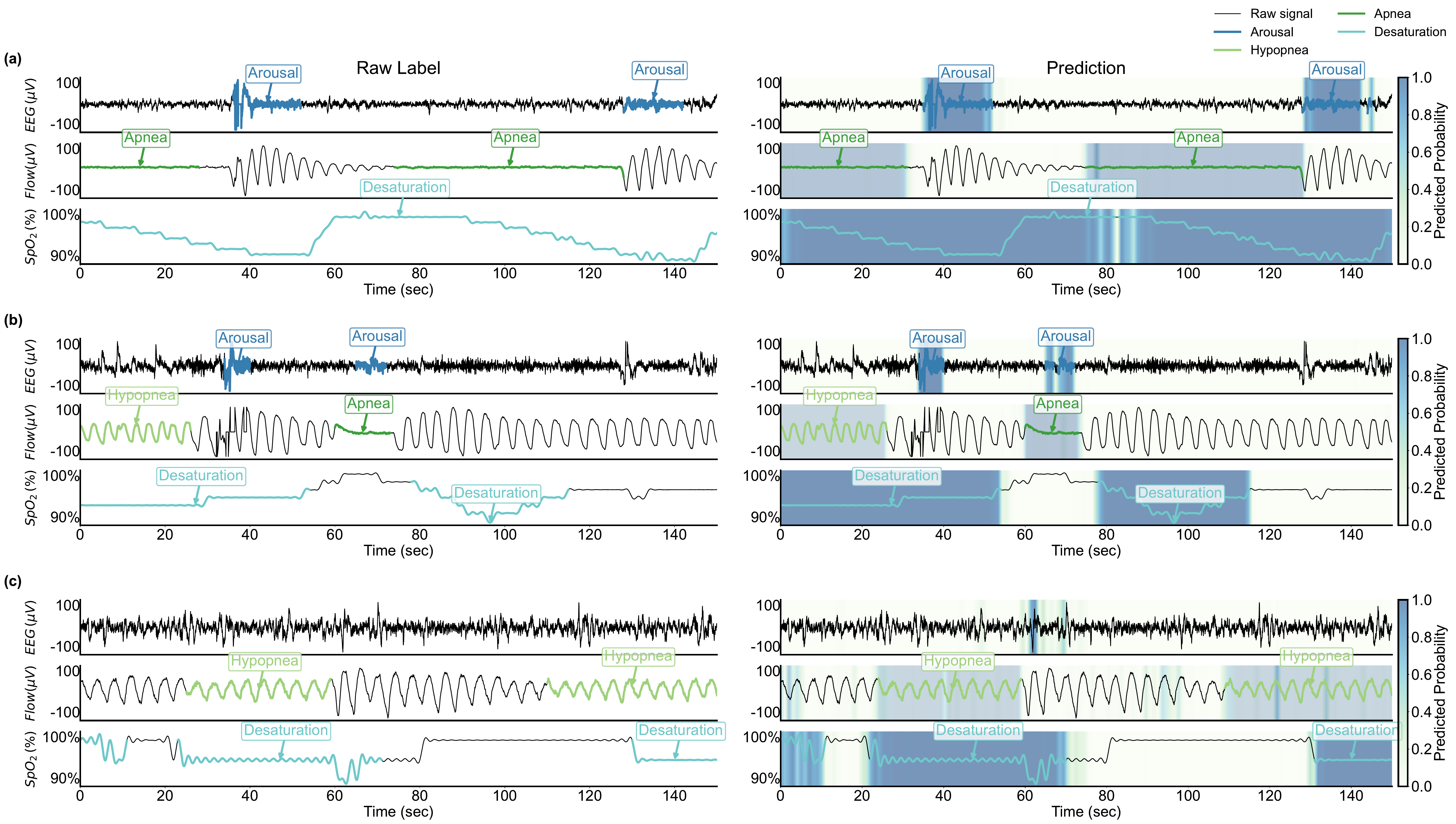}
\caption{Visualization of model predictions for three epoch conditions. The top panel represents an epoch containing both apnea and hypopnea events. The middle panel corresponds to an epoch with only apnea events, whereas the bottom panel shows an epoch with only hypopnea events. In all three cases, the model demonstrates reasonable predictive performance, accurately capturing the presence and timing of respiratory events.}\label{fig_research}
\end{figure*}

Figure \ref{fig_research} illustrates the performance of the proposed model in detecting OSA-related events across three epoch conditions. The first row depicts an epoch containing both apnea and hypopnea events; the model can successfully identify the events with a high degree of temporal alignment. The second row presents an epoch characterized solely by apnea events, and the model effectively captures these events. The third row displays an epoch containing only hypopnea events, and the model can accurately detect these events.

\subsection{Clinical OSA Severity Assessment}
\begin{figure*}[t]
    \centering
    \includegraphics[width=0.75\linewidth]{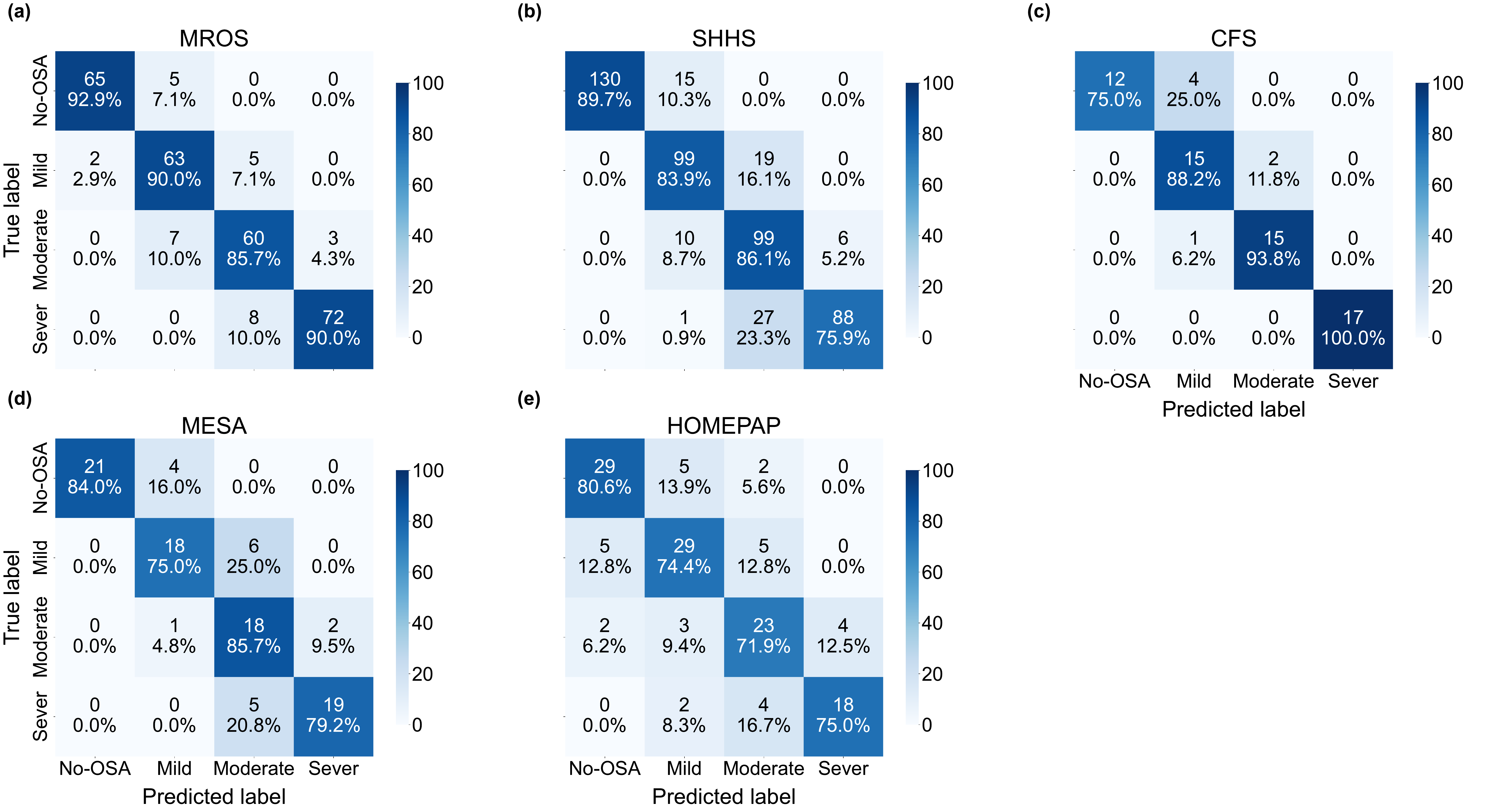}
    \caption{Confusion matrices for clinical OSA severity assessment: OSA severity classification, evaluating the model’s ability to categorize subjects into four OSA severity levels: Non-OSA ($AHI < 5$), Mild ($5 \leq AHI < 15$), Moderate (15 $\leq AHI < 30$), and Severe (AHI $\geq$ 30). Each subplot represents classification performance for a specific dataset: (a) MROS, (b) SHHS, (c) MESA, (d) CFS, and (e) HOMEPAP. The x-axis represents predicted severity categories, while the y-axis represents true severity levels. Darker blue shades indicate higher classification accuracy, with percentages displayed in each cell. Misclassifications primarily occur at the boundaries of adjacent severity levels, reflecting the challenge of differentiating borderline cases.}
    \label{fig_clinical}
\end{figure*}

With multimodal input (EEG, airflow, $SpO_2$), UT-OSANet provided continuous AHI estimation and classified OSA severity into four categories. Overall accuracy ranged from 76\% to 90\% and macro-F1 from 0.76 to 0.90 (Supplementary Table~S2). 
The model achieved 90\% and 84\% accuracy with the MROS and SHHS datasets, respectively, with strong classification performance across all OSA severity levels. The confusion matrices (Figure \ref{fig_clinical}(a-b)) show that most misclassifications occurred between mild and moderate OSA cases, which is expected given their overlapping characteristics in the AHI distribution. The model exhibited slightly lower accuracy (81\%) with the MESA dataset, primarily due to the increased misclassification rate between mild and moderate OSA (Figure \ref{fig_clinical}(c)), suggesting potential dataset-specific variability in model performance. The model achieves the highest classification accuracy for severe OSA (100\%) with the CFS dataset, indicating that the model is highly reliable in identifying severe cases (Figure \ref{fig_clinical}(e)). The model has the lowest accuracy (76\%) with the HOMEPAP dataset, likely due to dataset variability and differences in data acquisition protocols (Figure \ref{fig_clinical}(d)), with more misclassifications occurring between adjacent severity levels. 

\begin{figure*}[t]
    \centering
    \includegraphics[width=0.75\linewidth]{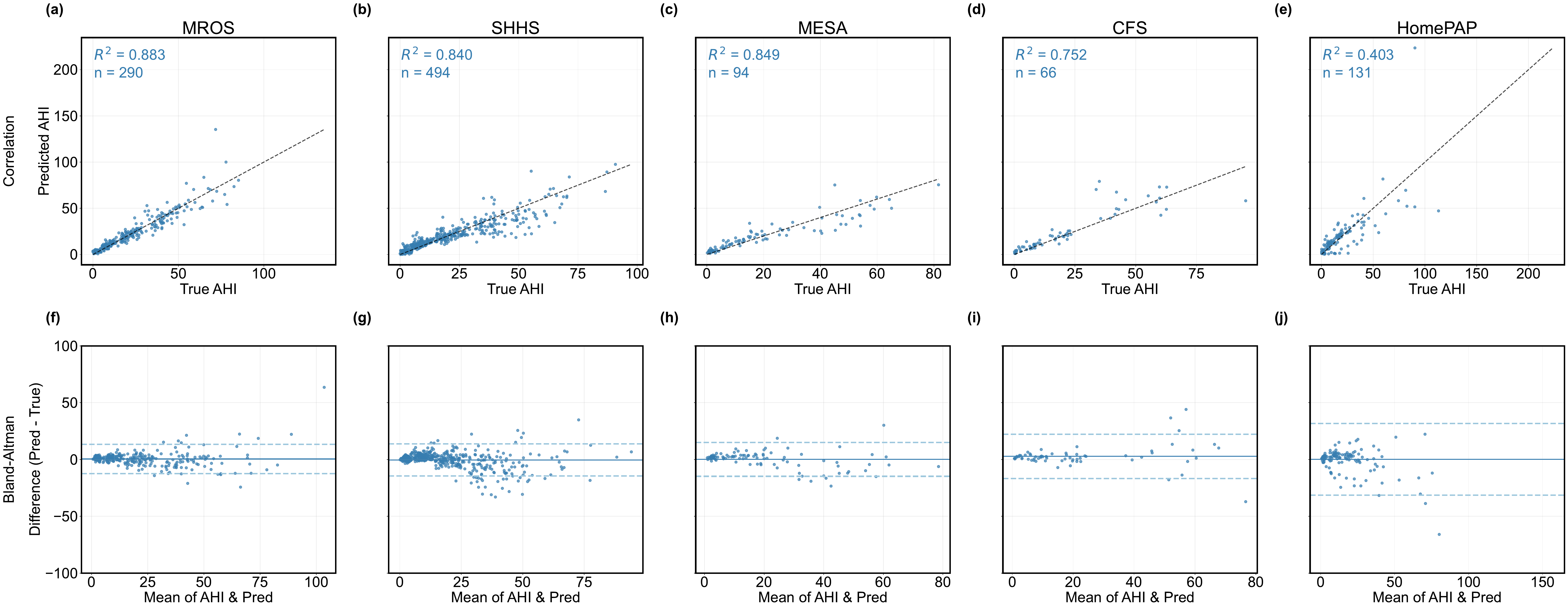}
    \caption{Correlation and Bland–Altman plots for homebase-OSA screening: AHI estimation across different datasets. The first row presents scatter plots comparing predicted AHI values with ground truth AHI values for (a) MROS, (b) SHHS, (c) MESA, (d) CFS, and (e) HOMEPAP. The diagonal dashed line represents the ideal 1:1 agreement. The second row displays the Bland–Altman plots, illustrating the difference between predicted and true AHI values against their mean. The red dashed lines indicate the 95\% limits of agreement, providing insights into systematic bias and variability in AHI estimation.}
    \label{fig_ahi}
\end{figure*}

Figure \ref{fig_ahi} presents scatter plots (top row) and Bland--Altman plots (bottom row) comparing predicted and reference AHI values across datasets. 
In the scatter plots, most data points are closely aligned with the identity line ($y = x$), reflecting strong correlations between predicted and true AHI values. 
Specifically, the model achieved high coefficients of determination ($R^2$) for the MROS ($R^2 = 0.883$, $n=290$), SHHS ($R^2 = 0.840$, $n=494$), and MESA ($R^2 = 0.849$, $n=94$) datasets, indicating reliable estimation across both community- and research-based cohorts. 
Performance on the CFS dataset was moderately strong ($R^2 = 0.752$, $n=66$), whereas the HOMEPAP dataset yielded a lower correlation ($R^2 = 0.403$, $n=131$), likely reflecting greater data heterogeneity and differences in acquisition protocols.

\subsection{Home-Based OSA Screening}\label{subsec1}
UT-OSANet was evaluated for home-based OSA screening using only EEG signals. Following AASM criteria \cite{berry2012respiratory}, an estimated AHI was computed by combining apnea and hypopnea events accompanied by either arousal or $SpO_2$ desaturation. Although this simplified estimate may deviate from PSG reference values, the model effectively distinguished $AHI < 15$ from $AHI \geq 15$, the threshold for moderate-to-severe OSA.

Across five datasets, UT-OSANet achieved accuracies ranging from 77\% (HOMEPAP) to 97\% (MROS), with consistently high recall and balanced precision (see Supplementary Table~S1). Figure~\ref{fig_homebase} illustrates confusion matrices across datasets, showing strong classification with most errors concentrated near the $AHI=15$ threshold.

\begin{figure*}[t]
\centering
\includegraphics[width=0.75\textwidth]{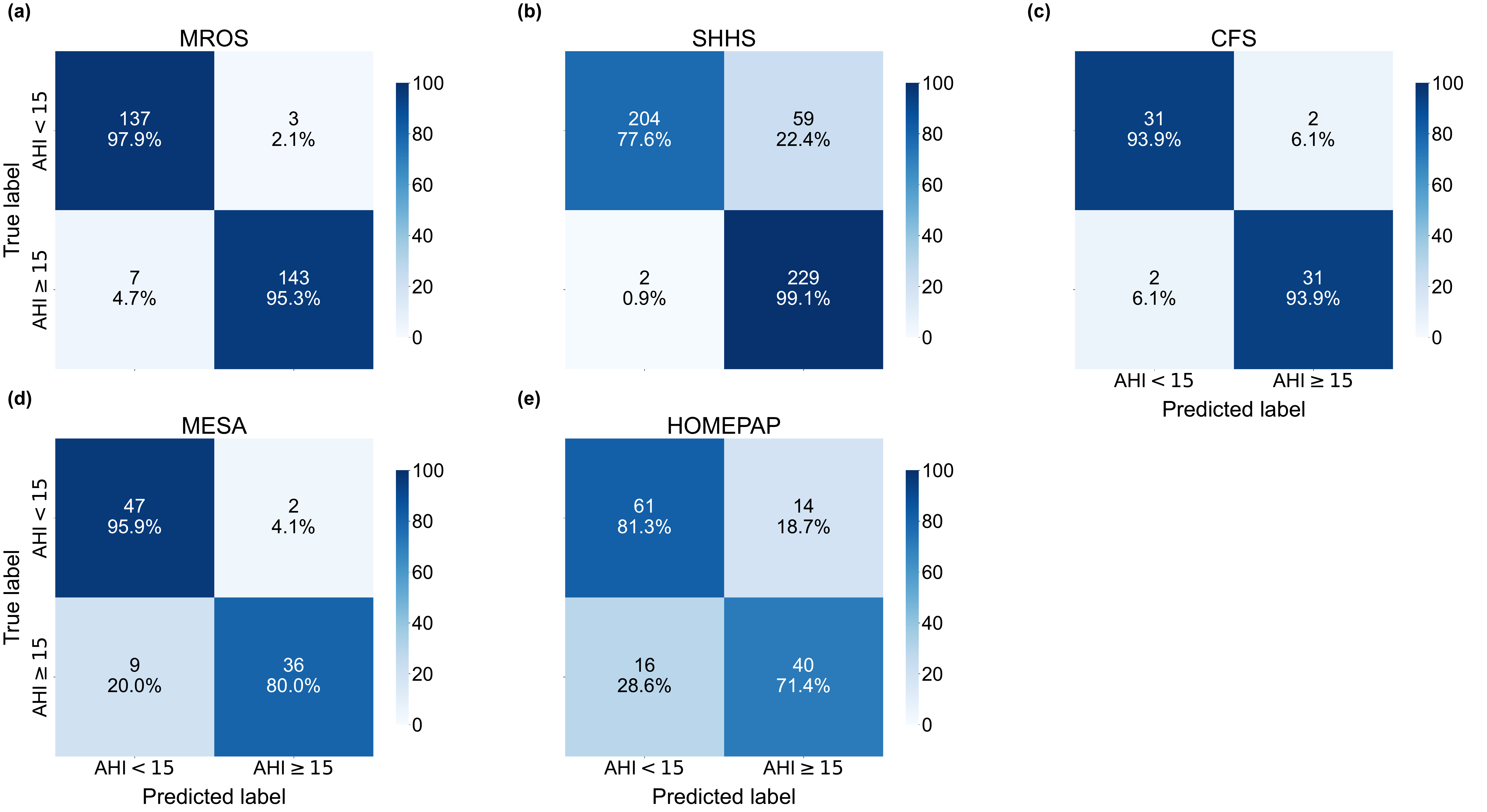}
\caption{Confusion matrices for home-base OSA screening, evaluating the classification of moderate-to-severe OSA (AHI $\geq$ 15) and below-moderate OSA ($AHI < 15$) across five independent datasets. Each subplot represents the classification performance for a specific dataset: (a) MROS, (b) SHHS, (c) MESA, (d) CFS, and (e) HOMEPAP. The x-axis represents the predicted labels, while the y-axis represents the true labels. Darker blue shades indicate higher classification accuracy, with percentages displayed in each cell. The majority of misclassifications occur near the AHI threshold, highlighting the model's performance in borderline cases.}\label{fig_homebase}
\end{figure*}


\section{DISCUSSION}
\begin{sidewaystable*}[t]
\centering
\caption{Comparison of representative OSA detection algorithms and their performance across different tasks}
\label{tab:OSA_comparison}
\begin{threeparttable}
\begin{tabular}{p{3cm} p{3cm} p{3cm} p{4cm} p{5cm} p{4cm}}
\toprule
\textbf{Study} & \textbf{Input modality} & \textbf{Algorithm/Model} & \textbf{AHI estimation and model performance} & \textbf{Event detection and model performance} & \textbf{Sample size} \\
\midrule
Jung et al. (2017) \cite{hwang2017real} & SpO$_2$ & Regression (morphometric features) & 2-class: AHI $\leq$ 30, $>$30; Sens = 0.97 & 1-event: Apnea; Acc = 0.91, $r$ = 0.71 & 230 PSG recordings (laboratory-based cohort) \\

Garde et al. (2015) \cite{garde2015pulse} & SpO$_2$ + PRV & Logistic regression & -- & 1-event: SpO$_2$ desaturation; AUC = 0.87, Sens $\approx$ 0.80--0.85 & 160 participants (smartphone-based pediatric cohort) \\

Levy et al. (2023) OxiNet \cite{levy2023deep} & SpO$_2$ & Deep learning (CNN) & 4-class: AHI $\leq$5, 5--15, 15--30, $\geq$30; ICC = 0.94--0.96, F$_{1,\mathrm{M}}$ = 0.75--0.84 & -- & 12,923 PSG recordings (multi-center cohort) \\

Berry et al. (2012) \cite{berry2012respiratory} & PAP airflow & Rule-based AED & -- & 3-event: Apnea, Hypopnea, Snoring; mean Sens = 0.58 & 115 device-derived recordings (PAP system dataset) \\

Koley et al. (2013) \cite{koley2013real} & Airflow & SVM (handcrafted features) & 3-class: AHI $\leq$5, 5--15, $>$15; Acc = 0.965 & 2-event: Apnea; Sens = 0.944; Hypopnea; Sens = 0.919 & 36 PSG recordings (limited dataset) \\

Van Steenkiste et al. (2020) \cite{van2020portable} & BioZ & LSTM & -- & 3-event: OSA, CSA, Hypopnea; mean Sens = 0.58 & 25 portable recordings (bioimpedance-based cohort) \\

Ben-Israel et al. (2012) \cite{ben2012obstructive} & Snoring & Feature-based & 2-class: AHI $\leq$20, $>$20; Acc = 0.83 & -- & 90 participants (snoring-sound dataset) \\

Huttunen et al. (2022) \cite{huttunen2022comparison} & PPG + SpO$_2$ + nasal pressure & Deep learning & 4-class: AHI $\leq$5, 5--15, 15--30, $\geq$30; ICC = 0.93--0.94 & 2-event: Apnea; Sens = 0.71; Hypopnea; Sens = 0.55 & 877 participants (multimodal signal dataset) \\

Yook et al. (2024) \cite{yook2024deep} & Airflow + SpO$_2$ + ECG + demographics & Xception & 4-class: AHI $\leq$5, 5--15, 15--30, $\geq$30; Acc = 0.89 & 2-event: Apnea; Sens = 0.97; Hypopnea; Sens = 0.92 & 923 participants (clinical cohort) \\

Rossi et al. (2023) \cite{rossi2023sleep} & Optical + air pressure + motion & Hybrid model & -- & 1-event: Apnea; Sens = 0.84 & 20 wearable recordings (chest-patch dataset) \\

Zahid et al. (2023) \cite{zahid2023msed} & Multimodal PSG & Multitask deep learning & 4-class: AHI $\leq$5, 5--15, 15--30, $\geq$30; $R^2$ = 0.77 & 3-event: Arousal; Sens = 0.67; Leg movement; Sens = 0.64; SDB; Sens = 0.53 & 2,653 PSG recordings (multimodal dataset) \\

\textbf{UT-OSANet (This study)} & EEG + airflow + SpO$_2$ (flexible) & Deep learning (multimodal) & 4-class: AHI $\leq$5, 5--15, 15--30, $\geq$30; F$_{1,\mathrm{M}}$ = 0.76--0.90 & 4-event: Apnea; Sens = 0.78--0.88; Hypopnea; Sens = 0.78--0.89; Desaturation; Sens = 0.79--0.89; Arousal; Sens = 0.81--0.90 & 9,021 PSG recordings (multi-cohort validation: MROS, SHHS, MESA, CFS; HOMEPAP as independent test) \\
\bottomrule
\end{tabular}
\end{threeparttable}
\end{sidewaystable*}

In recent years, numerous machine learning and deep learning methods have been proposed for automated OSA assessment based on specific physiological signals. These approaches—ranging from traditional regression and SVM classifiers to modern CNN and LSTM architectures—have achieved notable accuracy in detecting OSA presence using single modalities such as oximetry, airflow, or snoring \cite{ferreira2022enabling, duarte2023role}. However, as summarized in Table~\ref{tab:OSA_comparison}, most existing models remain limited in three critical aspects.
First, their diagnostic granularity is coarse: they primarily distinguish between OSA and non-OSA cases, with limited capability to characterize OSA severity or identify specific respiratory events.
Second, their input design is fixed to one or few modalities, making them difficult to apply consistently across diverse clinical and home environments where signal availability varies.
Third, these models are often developed on small, homogeneous cohorts, restricting their generalizability to broader populations.
Although a few multimodal frameworks (e.g., Huttunen et al., Yook et al., Zahid et al.) have attempted to combine complementary biosignals, their scalability and cross-dataset robustness remain inadequate.
In contrast, UT-OSANet was developed to overcome these limitations by integrating EEG, nasal airflow, and $SpO_2$ within a unified deep learning framework validated on 9,021 PSG recordings from five independent cohorts. Its flexible input configuration enables adaptive operation under varying modality conditions—from EEG-only home screening to comprehensive clinical diagnosis—achieving both fine-grained, event-level precision and strong generalization across datasets.

In the development of UT-OSANet, we selected the input modalities according to both the physiological complexity of sleep and the diagnostic requirements outlined by the AASM \cite{thornton2012aasm}. While PSG offers a wide array of data types--including EEG, electrocardiography (ECG), $SpO_2$ levels, airflow, and limb movements--not all signals contribute equally to the detection of core respiratory events \cite{rundo2019polysomnography}. To calculate the AHI, apnea, hypopnea, $SpO_2$ desaturation, and arousal events must be accurately identified \cite{berry2017aasm}; thus, we prioritized EEG, $SpO_2$ levels, and airflow data as the primary input modalities. These signals provide the most direct and interpretable physiological markers for respiratory event detection. Furthermore, UT-OSANet was explicitly designed as a framework that could be adapted to various modalities, allowing the input configuration to be flexibly adjusted according to the application scenario. This design ensures that the model maintains high diagnostic accuracy while maximizing its practical applicability across various scenarios, including home-based screening, clinical severity assessment, and research-grade event analysis.

The superior performance of UT-OSANet can be attributed to two key aspects: the hybrid U-Net and transformer architecture and the use of multimodal input features. The model structure, as shown in Figure \ref{fig_clinical}, is specifically designed for long-sequence, multimodal semantic segmentation. The U-Net module efficiently captures local spatial dependencies and
preserves fine-grained temporal features,
whereas the transformer module enables long-range temporal modeling, allowing the model to detect apnea, hypopnea, arousal, and oxygen desaturation events with high precision \cite{azad2024medical, gao2021utnet}. Another crucial factor is the integration of three physiological modalities: EEG, airflow, and blood oxygen saturation ($SpO_2$) levels. Each modality provides complementary information about OSA-related events: EEG captures arousal-related activity, airflow levels can be used to detect apnea and hypopnea events, and $SpO_2$ levels reflect oxygen desaturation patterns \cite{thornton2012aasm}. This multimodal approach significantly increases the model's ability to distinguish normal and pathological respiratory events, increasing the classification and segmentation accuracy \cite{yook2024deep, huttunen2022comparison, zahid2023msed}. By leveraging both an advanced deep learning architecture and rich physiological signals, UT-OSANet achieves robust and generalizable OSA detection, outperforming conventional methods that rely on single-modality features \cite{hwang2017real, garde2015pulse, levy2023deep, berry2012respiratory, koley2013real, van2020portable, ben2012obstructive}. The combination of effective feature extraction, temporal dependency modeling, and multi-signal fusion ensures that the model performs well across different datasets, demonstrating its potential for automated, high-precision OSA screening in both clinical and home-based environments.

\begin{figure*}[t]
    \centering
    \includegraphics[width=1\linewidth]{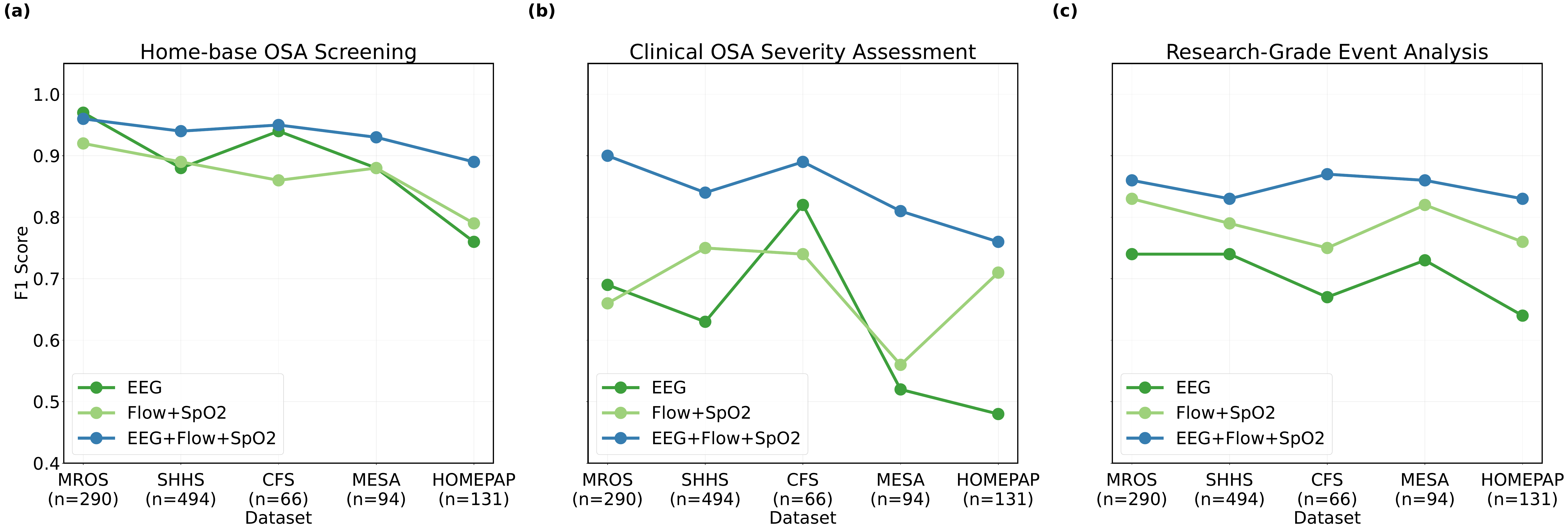}
    \caption{Home-base OSA screening: moderate and severe OSA Screening evaluates the ability to classify individuals into two categories: $AHI < 15$ vs. AHI $\geq$ 15, using the macro-F1 score.Clinical OSA Severity Assessment : OSA Classification assesses model performance in distinguishing four severity levels: No-OSA ($AHI < 5$), Mild (5 $\leq AHI < 15$), Moderate (15 $\leq AHI < 30$), and Severe (AHI $\geq$ 30). The macro-F1 score is reported to balance class disparities. Research-Grade Event Analysis: Event Detection focuses on identifying specific OSA-related events, including Apnea, Hypopnea, Arousal, and SpO2 Desaturation, where the macro-F1 score is computed across these four event classes.}
    \label{fig_baseline}
\end{figure*}
Home-based OSA screening presents unique challenges, primarily due to the limited availability of physiological signals and the necessity of user-friendly, portable hardware \cite{abd2024detection}. Conventional home sleep apnea tests (HSATs), such as pulse oximetry devices and wearable rings, predominantly rely on peripheral oxygen saturation ($SpO_2$) measurements \cite{park2024fda}. While these devices may be convenient to use, their diagnostic accuracy is limited, particularly in detecting apnea events without significant oxygen desaturation, which underscores the need for practical solutions \cite{duarte2023role}. In this study, we propose the use of single-channel EEG data combined with the UT-OSANet model for home-based OSA screening. As illustrated in the rightmost panel of Figure \ref{fig_homebase} (home-based OSA screening) and Supplementary Table~S4, we systematically evaluated model performance across three input configurations (EEG only, airflow + $SpO_2$, and EEG + airflow + $SpO_2$) and five independent test datasets, the results demonstrate that even with only EEG data as input, UT-OSANet achieves competitive--and often superior--F1 scores compared with those of the airflow + $SpO_2$ configuration in home-based OSA screening (Scenario 1), achieving F1 scores of 0.97 (MROS), 0.88 (SHHS), 0.94 (CFS), 0.88 (MESA), and 0.76 (HOMEPAP), outperforming the airflow + $SpO_2$ model with the MROS, CFS, and MESA datasets. Moreover, when all three modalities (EEG + airflow + $SpO_2$) are combined, the model consistently provides the best performance across datasets, with the F1 scores in scenario 1 reaching 0.96 (MROS), 0.94 (SHHS), 0.95 (CFS), 0.93 (MESA), and 0.89 (HOMEPAP). In contrast to $SpO_2$-based methods, EEG data can be used to identify arousal events and cortical activity linked to respiratory events \cite{goldstein2024polysomnography}. Since OSA is a sleep-related disorder, EEG plays a central role in sleep research by providing the primary physiological basis for sleep staging \cite{zhang2024review} and enabling the precise detection of arousal events \cite{giannakakis2019review} and cortical activity changes \cite{cohen2017does}. We demonstrated that even single-channel EEG data can be used for moderate-to-severe OSA screening and sleep assessment in home-based settings.

Compared with home-based screening, OSA assessment in clinical settings has distinct requirements and offers several opportunities. In clinical practice, sleep laboratories and hospitals have access to multimodal physiological signal data, including EEG, nasal airflow, peripheral oxygen saturation ($SpO_2$), ECG, and respiratory effort signals, enabling a detailed evaluation of OSA severity following standards such as those established by the AASM guidelines \cite{berry2017aasm}. Thus, robust event detection, including the precise identification of apnea, hypopnea, arousal, and desaturation events, is possible in clinical settings, supporting the classification of OSA severity into no, mild, moderate, and severe categories. In this study, we evaluated the performance of UT-OSANet in a clinical OSA assessment scenario (Scenario 2) with five independent datasets and systematically compared model performance across three input configurations: EEG only, airflow + $SpO_2$, and EEG + airflow + $SpO_2$. As summarized in Figure \ref{fig_frame} (clinical OSA assessment) and Supplementary Table~S4, the combined multimodal configuration (EEG + airflow + $SpO_2$) consistently showed the best performance across datasets, with F1 scores of 0.90 (MROS), 0.84 (SHHS), 0.89 (CFS), 0.81 (MESA), and 0.76 (HOMEPAP). While the EEG-only and airflow + $SpO_2$ configurations had moderately lower F1 scores--for example, the EEG-only configuration had F1 scores of only 0.69 (MROS), 0.63 (SHHS), 0.82 (CFS), 0.52 (MESA), and 0.48 (HOMEPAP)--the inclusion of all three signals significantly increased the classification accuracy, particularly in distinguishing mild, moderate, and severe OSA. These results highlight the value of leveraging multimodal physiological data in clinical scenarios, as the use of respiratory and neural signals enables more reliable and granular disease assessment. Importantly, EEG provides indispensable information on cortical arousal and sleep stage disruptions, whereas airflow and $SpO_2$ data provide critical insights into the mechanical and oxygenation aspects of respiratory events. Together, these complementary inputs allow UT-OSANet to precisely classify OSA severity, supporting detailed clinical decision-making and individualized treatment planning.

Research-grade OSA analysis is associated with even more demanding requirements on data richness and model interpretability, as such analyses are not limited to clinical diagnosis but also include detailed physiological mechanisms, comorbidities, and the relationships between respiratory events and broader health outcomes \cite{oksenberg2023duration}. In research settings, multimodal physiological signals, including EEG, nasal airflow, $SpO_2$, ECG, respiratory effort, and limb movement data, are typically available, enabling second-by-second or even finer temporal analysis of apnea, hypopnea, and desaturation events, cortical arousal, and leg movements. Researchers aim not only to detect the occurrence of these events but also to study their co-occurrence patterns, temporal dependencies, and relationships with diseases such as cardiovascular conditions, neurocognitive impairment, and metabolic syndromes \cite{randerath2025central, bouloukaki2024preserved, vanek2020obstructive, kerner2016obstructive}. In this study, we evaluated the performance of UT-OSANet in research-grade analyses (Scenario 3) with five independent datasets and compared model performance across three input configurations: EEG only, airflow + $SpO_2$, and EEG + airflow + $SpO_2$. As shown in Figure \ref{fig_baseline} and Supplementary Table~S4 (research-grade analysis scenario), the multimodal configuration (EEG + airflow + $SpO_2$) was consistently associated with the highest F1 scores, with values of 0.86 (MROS), 0.83 (SHHS), 0.87 (CFS), 0.86 (MESA), and 0.83 (HOMEPAP). While the EEG-only and airflow + $SpO_2$ configurations showed reasonable performance (for example, the EEG-only model had F1 scores of 0.74 (MROS), 0.74 (SHHS), 0.67 (CFS), 0.73 (MESA), and 0.64 (HOMEPAP)), the model integrating all modalities had significantly increased precision in identifying the timing and type of respiratory and arousal events. Critically, in addition to coarse summary metrics (e.g., the AHI), UT-OSANet provides fine-grained, time-resolved predictions of event occurrences and cross-event relationships. This capability is particularly valuable for analyses of the interplay between physiological disruptions and disease mechanisms, such as how repeated nocturnal hypoxemia events contribute to hypertension or how cortical arousal modulates autonomic nervous system responses \cite{shahrbabaki2023sleep}. By providing second-level temporal resolution and enabling multimodal event correlation analysis, UT-OSANet represents a powerful tool for advancing the mechanistic understanding of breathing in individuals with sleep disorders and the broader health impacts of such disorders.

\section{Conclusion}
In conclusion, UT-OSANet was trained and validated on multiple large-scale public datasets, achieving high accuracy across diverse OSA assessment scenarios. Its consistent performance in home, clinical, and research-grade analyses—using various combinations of EEG, airflow, and $SpO_2$ signals—demonstrates strong robustness and scalability. These results highlight UT-OSANet as a reliable, high-precision framework for population-scale OSA detection and characterization.

\bibliographystyle{IEEEtran}      
\bibliography{sn-bibliography}    
\end{document}